\documentclass[preprint2]{aastex6}
\newcommand\corot{\emph{CoRoT}}
\newcommand\kepler{\emph{{\it Kepler}}}
\newcommand\vsini{$v$\,sin\,$i$}

\newcommand\teff{$T_{\rm eff}$}

\newcommand\logg{log {\it g}}

\newcommand\Msun{\hbox{$M_{\odot}$}}  %Msun
\newcommand\Rsun{\hbox{$R_{\odot}$}}  %Rsun
\newcommand\kms{\hbox{km\,s$^{-1}$}}  %Rsun

\def\Mjup{\hbox{$M_\mathrm{Jup}$}} %Mjup
\def\Rjup{\hbox{$R_\mathrm{Jup}$}} %Rjup
\newcommand{\pper}{{$P_\mathrm{orb}$}}                 % Orbital Period
\begin{document}


%% Use \author, \affil, plus the \and command to format author and affiliation 
%% information.  If done correctly the peer review system will be able to
%% automatically put the author and affiliation information from the manuscript
%% and save the corresponding author the trouble of entering it by hand.
%%
%% The \affil should be used to document primary affiliations and the
%% \altaffil should be used for secondary affiliations, titles, or email.

\shortauthors{Grziwa, Gandolfi, Csizmadia et al.}
\shorttitle{The transiting planet K2-31\,b}

\title{K2-31b, a grazing transiting hot Jupiter\\ on an 1.26-day orbit around a bright G7V star.}

%% Authors with the same affiliation can be grouped in a single

\author{Sascha~Grziwa\altaffilmark{1},
Davide~Gandolfi\altaffilmark{2,3}, 
Szilard~Csizmadia\altaffilmark{4}, 
Malcolm~Fridlund\altaffilmark{5,6},
Hannu~Parviainen\altaffilmark{7},
Hans\,J.~Deeg\altaffilmark{8,9}, 
Juan~Cabrera\altaffilmark{4}, 
Amanda\,A.~Djupvik\altaffilmark{10},
Simon~Albrecht\altaffilmark{11},
Enric\,B.~Palle\altaffilmark{8,9},
Martin~P\"atzold\altaffilmark{1},
Victor\,J.\ S.\ B\'ejar\altaffilmark{8,9},
Jorge~Prieto-Arranz\altaffilmark{8,9},
Philipp~Eigm\"uller\altaffilmark{4}, 
Anders Erikson\altaffilmark{4}, 
Johan\,P.\,U.~Fynbo\altaffilmark{12},
Eike~W.~Guenther\altaffilmark{13},
Artie~P.~Hatzes\altaffilmark{13}, 
Amanda~Kiilerich\altaffilmark{11}, 
Judith~Korth\altaffilmark{1},
Teet~Kuutma\altaffilmark{10},
Pilar~Monta\~n\'es-Rodr\'iguez\altaffilmark{8,9},
David~Nespral\altaffilmark{8,9},
Grzegorz~Nowak\altaffilmark{8,9},
Heike~Rauer\altaffilmark{4,14} 
Joonas~Saario\altaffilmark{10},
Daniel~Sebastian\altaffilmark{13},
Ditte~Slumstrup\altaffilmark{10,11}
}

\altaffiltext{1}{Rheinisches Institut f\"ur Umweltforschung an der Universit\"at zu K\"oln, Aachener Strasse 209, 50931 K\"oln, Germany}
\altaffiltext{2}{Dipartimento di Fisica, Universit\'a di Torino, via P. Giuria 1, 10125 Torino, Italy; oscar.barraganvil@edu.unito.it}
\altaffiltext{3}{Landessternwarte K\"onigstuhl, Zentrum f\"ur Astronomie der Universit\"at Heidelberg, K\"onigstuhl 12, 69117 Heidelberg, Germany}
\altaffiltext{4}{Institute of Planetary Research, German Aerospace Center, Rutherfordstrasse 2, 12489 Berlin, Germany}
\altaffiltext{5}{Leiden Observatory, University of Leiden, PO Box 9513, 2300 RA, Leiden, The Netherlands}
\altaffiltext{6}{Department of Earth and Space Sciences, Chalmers University of Technology, Onsala Space Observatory, 439 92 Onsala, Sweden}
\altaffiltext{7}{Department of Physics, University of Oxford, Oxford, OX1 3RH, United Kingdom}
\altaffiltext{8}{Instituto de Astrof\'isica de Canarias, 38205 La Laguna, Tenerife, Spain}
\altaffiltext{9}{Departamento de Astrof\'isica, Universidad de La Laguna, 38206 La Laguna, Spain}
\altaffiltext{10}{Nordic Optical Telescope, Apartado 474, 38700, Santa Cruz de La Palma, Spain}
\altaffiltext{11}{Stellar Astrophysics Centre, Department of Physics and Astronomy, $\AA$rhus Uni., Ny Munkegade 120, DK-8000 $\AA$rhus C, Denmark}
\altaffiltext{12}{Dark Cosmology Centre, Niels Bohr Institute, Copenhagen University, Juliane Maries Vej 30, 2100 Copenhagen O, Denmark}
\altaffiltext{13}{Th\"uringer Landessternwarte Tautenburg, Sternwarte 5, D-07778 Tautenberg, Germany}
\altaffiltext{14}{Center for Astronomy and Astrophysics, TU Berlin, Hardenbergstr. 36, 10623 Berlin, Germany}
\altaffiltext{15}{Departamento de F\'isica Aplicada, Universidad de Cantabria Avenida de los Castros s/n, E-39005 Santander, Spain}

\begin{abstract}
We report the discovery of K2-31b, the first confirmed transiting hot Jupiter detected by the K2 space mission. We combined K2 photometry with FastCam lucky imaging and FIES and HARPS high-resolution spectroscopy to confirm the planetary nature of the transiting object and derived the system parameters. K2-31b is a 1.8-Jupiter-mass planet on an 1.26-day-orbit around a G7\,V star ($M_\star=0.91$~\Msun, $R_\star=0.78$~\Rsun). The planetary radius is poorly constrained (0.7$<$$R_\mathrm{p}$$<$1.4~\Rjup)\footnote{The short transit-duration of ~55 min excludes low-latitude transits, whose planet-sizes can be better constrained.}, owing to the grazing transit and the low sampling rate of the K2 photometry\footnote{In a very rare case at the lower limit of the planetary radius the transit is not ``grazing'' but 'nearly grazing'. We stay with the term ``grazing'' throughout the text for  simplicity and because ``nearly grazing'' transits produce the same transit signal.}. 
\end{abstract}




\section{Introduction} 
\label{sec:intro}

The advent of space-based transit surveys such as \corot\ \citep{baglin_2006b} and  \kepler\ \citep{borucki_2010} has provided a major breakthrough for exoplanetary science, in terms of both the number of discoveries and the precision of measured exoplanet radii. Unfortunately, the second reaction wheel of the \kepler\ spacecraft failed in May 2013 affecting the pointing precision of the telescope. The mission -- now called K2 -- has been resumed in 2014 by adopting a different observing strategy that uses only two reaction wheels and thrusters to control the pointing of the telescope \citep{howell_2014}. The capability of K2 to detect small transiting planets in short period orbits around relatively bright stars has recently been shown \citep[e.g.,][]{vander1_2015}.

Although much interest is focused on finding Earth-size planets because they are central to astrobiology, hot Jupiters still remain important targets for investigating migration mechanisms and star-planet tidal interactions. Only 21 hot Jupiters ($>0.3 \Mjup$) on planetary orbits smaller than 1.3-days are confirmed to date (August 2016; see \url{exoplanet.eu}).

We report here the discovery of \object{K2-31b}, a short-period giant planet transiting a V=10.8~mag star observed by the K2 mission. We combined the K2 photometry with ground-based follow-up observations to confirm the system and derive its fundamental parameters. We have also shown that grazing transits can entail a correlation between the relative stellar radius $R_p/R_*$ and the orbital inclination $i$ \footnote{In this work the orbital inclination describes the angle beetween the normal vector perpendicular to the orbital plane of the planet and the line of sight from the observer point of view. The smallest inclination of $=90^{\circ}$ means an edge-on observation of the transiting planet in front of the star.} when estimating the system parameters. This paper is the collaborative work of two teams dedicated to the detection, analysis and follow-up of K2 targets, namely the Kepler Exoplanet Science Team (KEST) and the ESPRINT team \citep{Sanchis-Ojeda2015}.
%To date only 12 hot Jupiters (see \url{exoplanet.eu}) with a period $p\le 1.26$ days were detected and less are confirmed.   

\section{K2 photometry} 
\label{sec:lcurve}

\begin{table}
\centering
\caption{Host star identifiers, equatorial coordinates, magnitude, and physical parameters.}
\label{Star-Param}
%\scalebox{0.9}{
\begin{tabular}{lr}
   \hline
   \hline
   \noalign{\smallskip}                
   \multicolumn{1}{l}{\emph{Main identifiers}} \\
   \hline
   \noalign{\smallskip}                
     EPIC            & 204129699 (K2-31)       \\
     TYC             & 6794-471-1        \\
     2MASS~ID        & 16214578-2332520  \\
  \noalign{\smallskip}                
  \multicolumn{1}{l}{\emph{Equatorial coordinates and magnitude}}  \\
  \hline
  \noalign{\smallskip}                
     RA \,(J2000)      & $16^h\,21^m\,45\fs801$             \\
     Dec\,(J2000)      & $-23\degr\,32\arcmin\,52\farcs32$  \\
     V\,[mag]          & $10.8$                             \\
  \noalign{\smallskip}
  \multicolumn{1}{l}{\emph{Stellar parameters}} \\
  \hline
  \noalign{\smallskip}
      Star mass [\Msun]                                    & $0.91\pm0.06$   \\ % $0.86\pm0.04$
      Star radius [\Rsun]                                  & $0.78\pm0.07$   \\ % $0.78\pm0.06$
      Effective Temperature \teff~[K]                      & $5280\pm  70$   \\
      Surface gravity \logg~(cgs)                          & $4.60\pm0.07$   \\
      Metallicity [Fe/H] [dex]                             & $0.08\pm0.05$   \\
      Microturbulent velocity $v_ {\mathrm{micro}}$~[\kms] & $0.9 \pm 0.1$   \\
      Macroturbulent velocity $v_ {\mathrm{macro}}$~[\kms] & $2.2 \pm 0.5$   \\            
      Rotational velocity \vsini~[\kms]                    & $2.6 \pm 0.5$   \\         
      Rotational period $P_\mathrm{rot}$~[day]             & $18.38\pm0.06$  \\
%      Age~[Gyr]                                            & $ 1 \pm -- $    \\
%      Distance~[pc]                                        & $ -- \pm -- $   \\
%      Extinction $A_\mathrm{v}$~[mag]                      & $ -- \pm -- $   \\ 
      Spectral type                                        & \text{G}7\,\text{V} \\
    \noalign{\smallskip}
    \hline
\end{tabular}
%}
\end{table}

\begin{figure}[t] 
\begin{center}
\resizebox{\hsize}{!}{\includegraphics[angle=0]{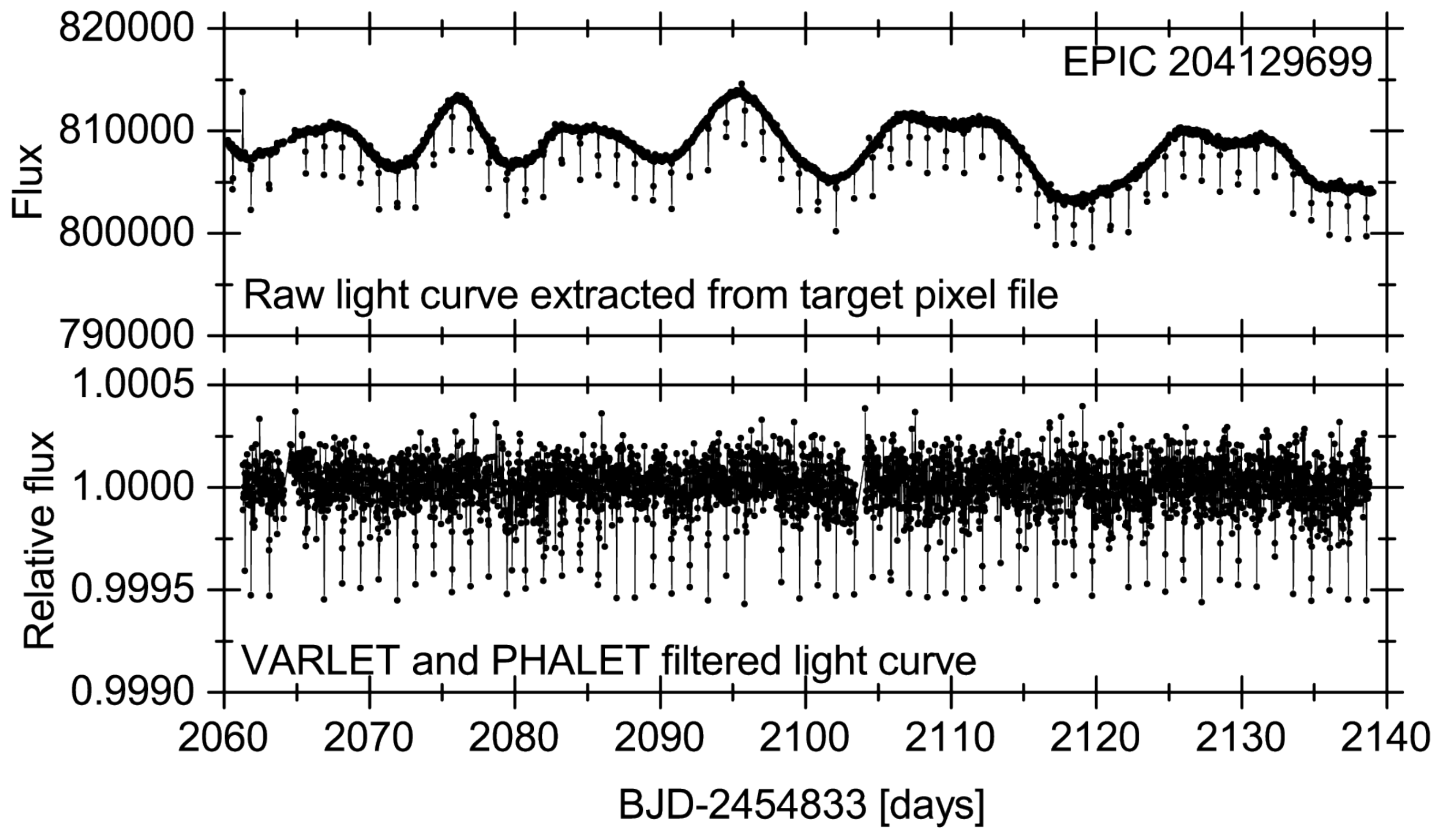}}
\caption{\emph{Upper panel}: K2-31's light curve extracted from the target pixel files. \emph{Lower panel}: Light curve filtered with VARLET and PHALET.}
\label{lc}
\end{center}
\end{figure}

The star \object{K2-31} (Table \ref{Star-Param}) was observed in the long cadence mode (T$_\mathrm{exp}$$\sim$29 min) by K2 for about 80 consecutive days as part of Campaign 2 (23 August - 13 November 2014). In contrast to the original Kepler mission, K2 provides only target pixel files and not ``ready-to-use'' detrended light curves. The target pixel files contain images with the calibrated flux, the subtracted background, and the removed cosmic rays. The light curve was automatically extracted by using a software developed by the Rhenish Institut for Environmental Research, department of planetary research (RIU-PF). The software detects the target stars in the pixel files, calculates the stellar masks, compensates the translation of the targets on the CCD by shifting the masks, integrates the flux within the stellar masks, and creates a light curve by listing the integrated flux from each pixel file as a function of time (Fig.~\ref{lc}, upper panel). Flux variations caused by the rotational drift of the spacecraft and the periodic pointing corrections of the thrusters are corrected using a technique similar to those used by \citet{vander_2014} and \citet{Lund2015}.

The K2 light curves of Campaign 2 were independently processed and searched for planetary transits by two teams within our collaboration, namely, the DLR in Berlin and the RIU-PF in Cologne, using two different software packages (\texttt{DST}, \citealt{cabrera_2012}; \texttt{EXOTRANS}, \citealt{grziwa_2012}). Prior to the transit search, the Cologne team filtered the light curves using the wavelet based filter \texttt{VARLET} and \texttt{PHALET} \citep{grziwa_2016} to reduce stellar variabilities and instrument systematics (Fig.~\ref{lc}, lower panel). 

\begin{figure}[t] 
\begin{center}
\resizebox{\hsize}{!}{\includegraphics[angle=0]{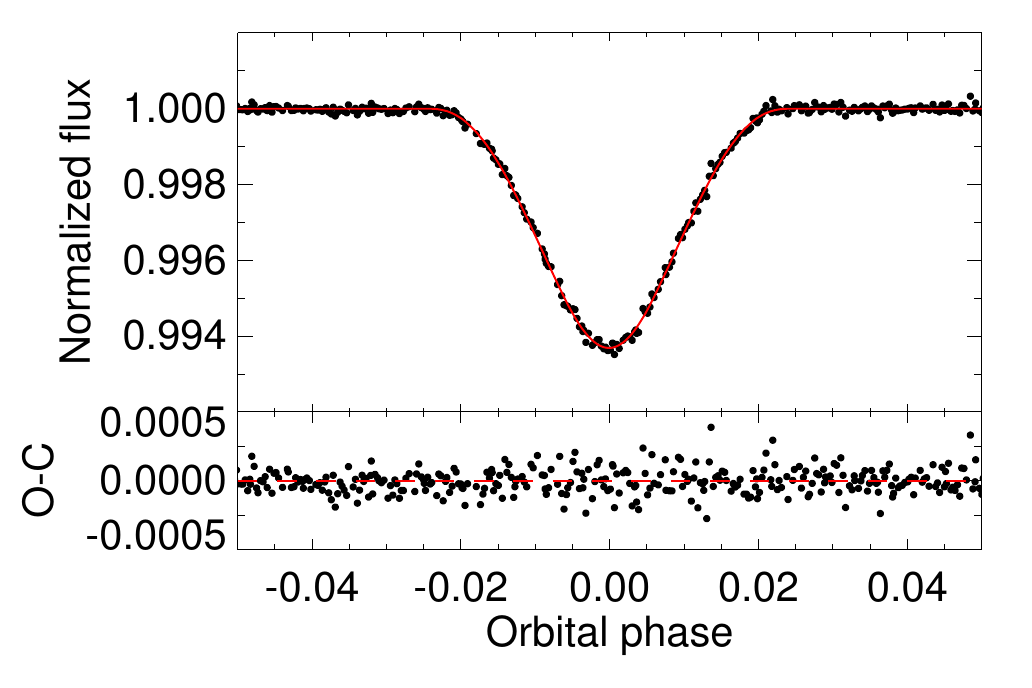}}
\caption{\emph{Upper panel}: Phase-folded transit light curve of K2-31b and an exemplary best-fitting transit model (red). \emph{Lower panel}: Residuals to the transit fit.}
\label{transit}
\end{center}
\end{figure}

A transit-like signal with a depth of 0.6\,\% occurring every $\sim$1.26 days was detected by both teams in the light curve of K2-31 (Fig.~\ref{transit}). Since the transit is V-shaped it was suspected that it might be caused by an eclipsing binary whose light contaminates the mask of the target star. The candidate was not discarded though, because grazing transiting planets exhibit also V-shaped signals. Since the star is relatively bright and first reconnaissance observations can be carried out quickly, the candidate was included in the list for follow-up observations. The confirmation of the planetary nature of this candidate came very promptly as described in Sect.~\ref{FU}.

The unfiltered light curve of K2-31 shows also quasi-periodic flux variations with a peak-to-peak amplitude of $\sim$1\,\% (Fig.~\ref{lc}). The flux variability is very likely caused by magnetic active regions moving in and out of sight as the star rotates. The stellar rotation period was inferred to be $P_\mathrm{rot}=18.38\pm0.06$\,days using the autocorrelation method described in \citet{McQuillan2014}.

\section{Ground-based follow-up observations}
\label{FU}

Follow-up observations were performed in order to exclude the false positive scenario \citep{Brown2003} and to derive the mass of the planet. High resolution imaging of K2-31 was obtained on 15 April 2015 using the FastCam lucky imaging camera at the 1.5-m Carlos Sanchez Telescope in Tenerife. Ten data cubes of 1\,000 images were taken, each with an exposure time of 50\,ms. The data were reduced following \citet{jodar_2013}. The frames were shifted and co-added to produce a final, high SNR, high resolution image. K2-31 appears to be an isolated object. Potential visible companions with $3\le\Delta$mag\,$\le 6$ can be excluded up to $3\arcsec$ from the target star. Companions with $\Delta$mag\,$\le 3$ can be discarded as close as $0.3\arcsec$.

The radial velocity (RV) follow-up of K2-31 started in May 2015 with the FIES spectrograph \citep{Telting2014} mounted at the 2.56-m Nordic Optical Telescope (NOT) of Roque de los Muchachos Observatory (La Palma, Spain). Six high-resolution spectra (R=67\,000) at an exposure time to 15--20 minutes were aquired, which resulted in a SNR of 30--38 per pixel at 5500\,\AA\ in the extracted spectra. The RV drift of the instrument was traced by taking long-exposure ($T_\mathrm{exp}$$\approx$30\,sec) ThAr spectra right before and after each observation following the observing strategy described by \citet{Gandolfi2015}. The data were reduced using standard procedures and RV measurements were extracted via multi-order cross-correlation with the RV standard star \object{HD\,182572}, observed with the same instrument set-up as the target object.

Three additional spectra were collected in June 2015 with the HARPS spectrograph \citep{Mayor03} mounted at the 3.6-m ESO telescope (La Silla, Chile) as part of the ESO program 095.C-0718(A). The exposure time was 20~minutes resulting in a SNR of 36--56 per pixel at 5500\,\AA\ in the extracted spectra. Radial velocities were obtained by multi-order cross-correlation with a numerical mask using the HARPS pipeline.

\begin{figure}[t]
\begin{center}
\resizebox{\hsize}{!}{\includegraphics[angle=0]{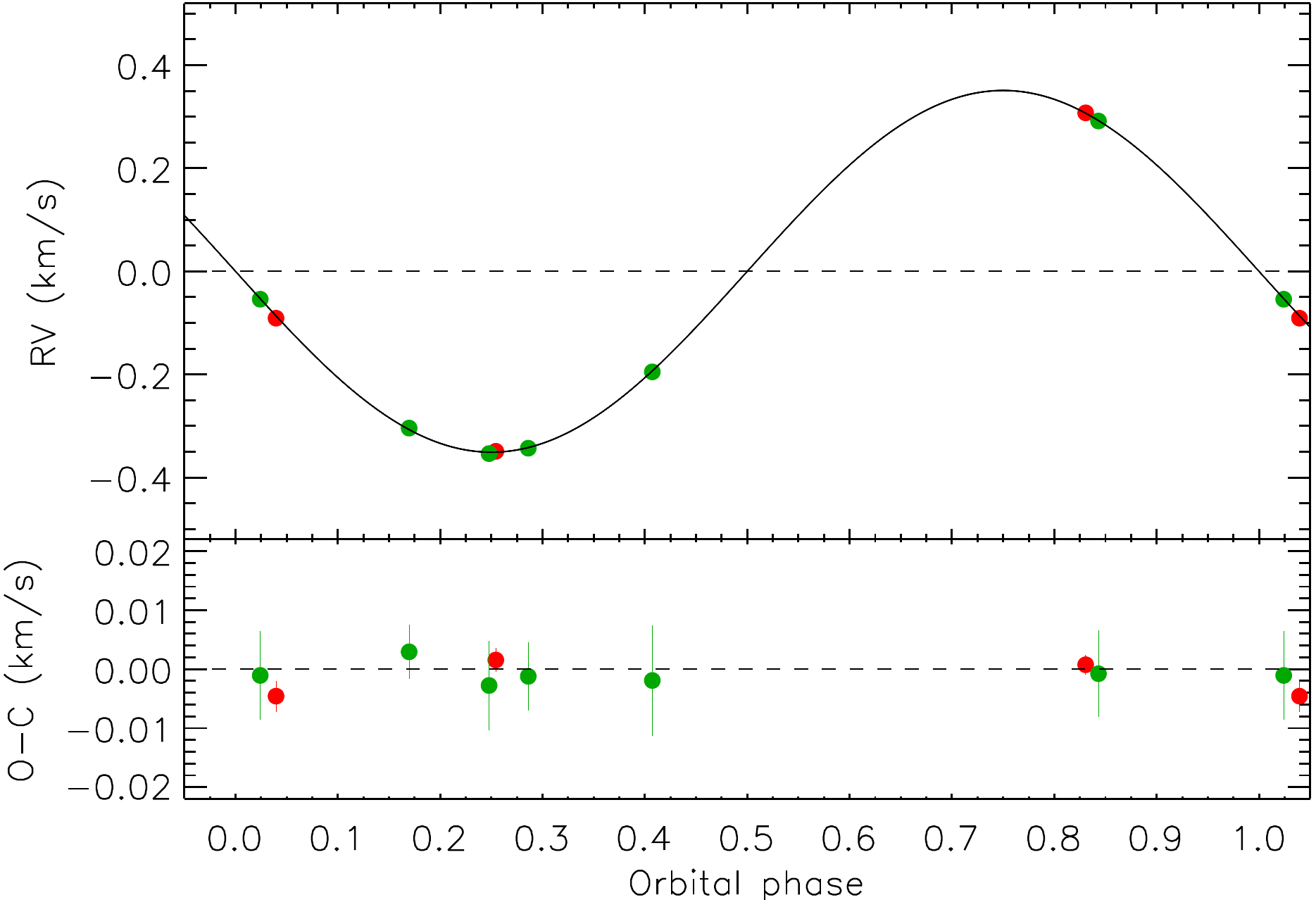}}
\caption{\emph{Upper panel}. Phase-folded FIES (green points) and HARPS (red points) RVs of K2-31 and Keplerian fit to the data. \emph{Lower panel}. Radial velocity residuals.}
\label{RV-Curve}
\end{center}
\end{figure}

The FIES and HARPS RV measurements are listed in Table~\ref{RV-Table}, along with the error bars, barycentric Julian dates in barycentric dynamical time, and the cross-correlation function (CCF) bisector spans. The two sets of RV measurements were fit to a Keplerian model, as described in Sect\,\ref{Modeling}. Figure~\ref{RV-Curve} shows the RV measurements -- after subtracting the RV offset between the two instruments and the systemic velocity -- phase-folded to the orbital period (upper panel), along with the RV residuals (lower panel). No significant correlation between the CCF bisector spans and the RVs was found indicating that the Doppler shifts observed in K2-31 are induced by the orbital motion of the planet rather than stellar activity or a blended eclipsing binary.

\begin{table}[t]
  %\centering 
  \caption{FIES and HARPS RVs of K2-31.}
  \label{RV-Table}
\begin{tabular}{cccccl}
  \hline
  \hline
  \noalign{\smallskip}                
BJD$_\mathrm{TDB}$ &    RV    & $\sigma_{\mathrm RV}$ &   Bis.    & Inst.  \\
($-$ 2\,450\,000)  &   \kms   &    \kms               &   \kms    &        \\
  \noalign{\smallskip}                
  \hline
  \noalign{\smallskip}                
7165.606212       & $-$5.2971 &       0.0076          & $-$0.0359 & FIES \\
7242.383219       & $-$5.2867 &       0.0058          & $-$0.0274 & FIES \\
7249.373260       & $-$4.6520 &       0.0073          & $-$0.0307 & FIES \\
7253.374466       & $-$4.9977 &       0.0075          & $-$0.0448 & FIES \\
7256.372026       & $-$5.1386 &       0.0094          & $-$0.0410 & FIES \\
7262.362392       & $-$5.2547 &       0.0046          & $-$0.0364 & FIES \\
7185.740109       & $-$5.1003 &       0.0014          & $-$0.0470 & HARPS \\
7186.727818       & $-$4.8421 &       0.0020          & $-$0.0393 & HARPS \\
7187.722887       & $-$4.4438 &       0.0011          & $-$0.0385 & HARPS \\  
  \noalign{\smallskip}                
  \hline
\end{tabular}
\tablecomments{Barycentric Julian dates are given in barycentric dynamical time. Radial velocity uncertainties, CCF bisector spans, and spectrographs are listed in the last three columns.}
\end{table}

\section{Fundamental parameters of the host star}

Two independent analyses were performed to estimate the spectroscopic parameters of K2-31 from the co-added FIES and HARPS spectra. The first method fits the observed spectra to a grid of theoretical models from \citet{Castelli2004}, \citet{Coelho2005}, and \citet{Gustafsson2008}, using spectral features that are sensitive to different parameters. The calibration equations for Sun-like stars from \citet{Bruntt2010} and \citet{Doyle2014} were adopted to determine the microturbulent $v_ {\mathrm{micro}}$ and macroturbulent $v_{\mathrm{macro}}$ velocities, respectively. The projected stellar rotational velocity $\vsini = 2.5 \pm 0.5$ km/s was measured by fitting the profile of several clean and unblended metal lines. The second method relies on the use of the spectral analysis package Spectroscopy Made Easy (\texttt{SME}) \citep[version 4.43][]{Valenti1996,Valenti2005}. \texttt{SME} calculates, for a set of given stellar parameters, synthetic spectra of stars are used to fit and observed high-resolution spectra using a $\chi^2$ minimization procedure. These two analyses provided consistent results well within the error bars, regardless of the method and spectrum used. The final adopted values are listed in Table~\ref{Star-Param}. A effective stellar temperature \teff$ = 5280 \pm 70 K$, a surface gravity log$g=4.60 \pm 0.07 $ and a metallicity $[Fe/H]=0.08 \pm 0.05$ was calculated. Using the \citet{Straizys1981} calibration scale for dwarf stars, the effective temperature of K2-31 defines the spectral type of the host star as G7\,V. A stellar mass of $0.91 \pm 0.06 \Msun$ and a stellar radius of $0.78 \pm 0.07 \Rsun$ were determined using the empirical calibrations from \citet{Torres2010}. Also \citet{Dai_2016} doppler monitored K2-31. They calculated slightly higher values for temperature (\teff$ =5412 \pm 34 K$), log g ($4.44 \pm 0.05$) and also higher values for metallicity ($0.20 \pm 0.05$). This results in a slightly different star model with a higher mass ($1.000 \pm 0.064 \Msun$) compared to our values but is consistent inside the error bars. They also calculated a higher stellar radius ($0.986 \pm 0.070 \Rsun$) compared to our values ($0.78 \pm 0.07 \Rsun$).

\section{Data modeling}
\label{Modeling}

By analyzing our FIES and HARPS measurements as described in section \ref{FU} we calculated a planetary mass of $1.774 \pm 0.079 \Mjup$ for K2-31b. In \citet{Dai_2016} a planetary mass of $1.856 \pm 0.084 \Mjup$ was calculated using their own PFS and TRES RV data. They also combined their PFS and TRES measurements with our FIES and HARPS data getting nearly the same results with a slightly smaller error ($1.857 \pm 0.081 \Mjup$). Nevertheless all three mass determinations are consistent inside the error bars and reveal the planetary nature of this hot Jupiter.\\
Unfortunately retrieving the system parameters from light curve analysis was more problematic.
The grazing transit of K2-31b implies that the light curve modeling alone cannot provide a unique best-fitting solution \citep{muller_2013,csizmadia_2013}. A correlation exists between the planet-radius ratio $R_\mathrm{p}/R_*$, the impact parameter $b$ (or inclination), and the semi-major-axis ratio $a_\mathrm{p}/R_*$. Constraints are thus needed to reduce the degeneracy of the parameter space. The impact parameter must be $b\gtrsim0.9$ following \citet{csizmadia_2015}. The estimate of the lower limit for the impact parameter uses the length of the transit and the estimated stellar mass and radius. The orbital period, the spectroscopically derived stellar mass and radius, and Kepler's third law, constrain the scaled semi-major axis to $a_\mathrm{p}/R_{\star}=6.05\pm0.50$, assuming a circular orbit\footnote{Although the RVs are better fitted by a 0.012 eccentricity, there is a $\sim$60\,\% probability that the best-fitting eccentric solution arises by chance if the orbit is actually circular \citep{Lucy1971}.}. Five different software packages were used for the transit fit based on the transit model by \citet{mandel_2002}. \texttt{EXOFAST} \citep{Eastman2013} and \texttt{PyTransit} \citep{Parviainen2015} were used to perform a simultaneous fit to the RV and K2 data and to obtain an estimate of the model posterior distribution using Markov chain Monte Carlo simulations. The Transit Light Curve Modeling (\texttt{TLCM}) code \citep{csizmadia_2015}, Transit Analysis Package (\texttt{TAP}) \citep{gazak_2012} and \texttt{UTM/UFIT} \citep{deeg_2014} were used to model the K2 data only.

\texttt{EXOFAST}, \texttt{PyTransit}, and \texttt{TAP} accounts for the K2 low acquisition rate by calculating the transit model for 8--10 sub-samples per long cadence data point. In the case of \texttt{TLCM}, the transit model was calculated for 33 equidistantly distributed points inside a long cadence measurement and then integrated using a Simpson-integrator. All five software packages showed the same result, which is listed in Table~\ref{System-Param}. The correlation between the impact parameter and the radius ratio can not be resolved although as many parameters as possible were constrained.

\begin{figure}[t]
\begin{center}
\resizebox{\hsize}{!}{\includegraphics[angle=0]{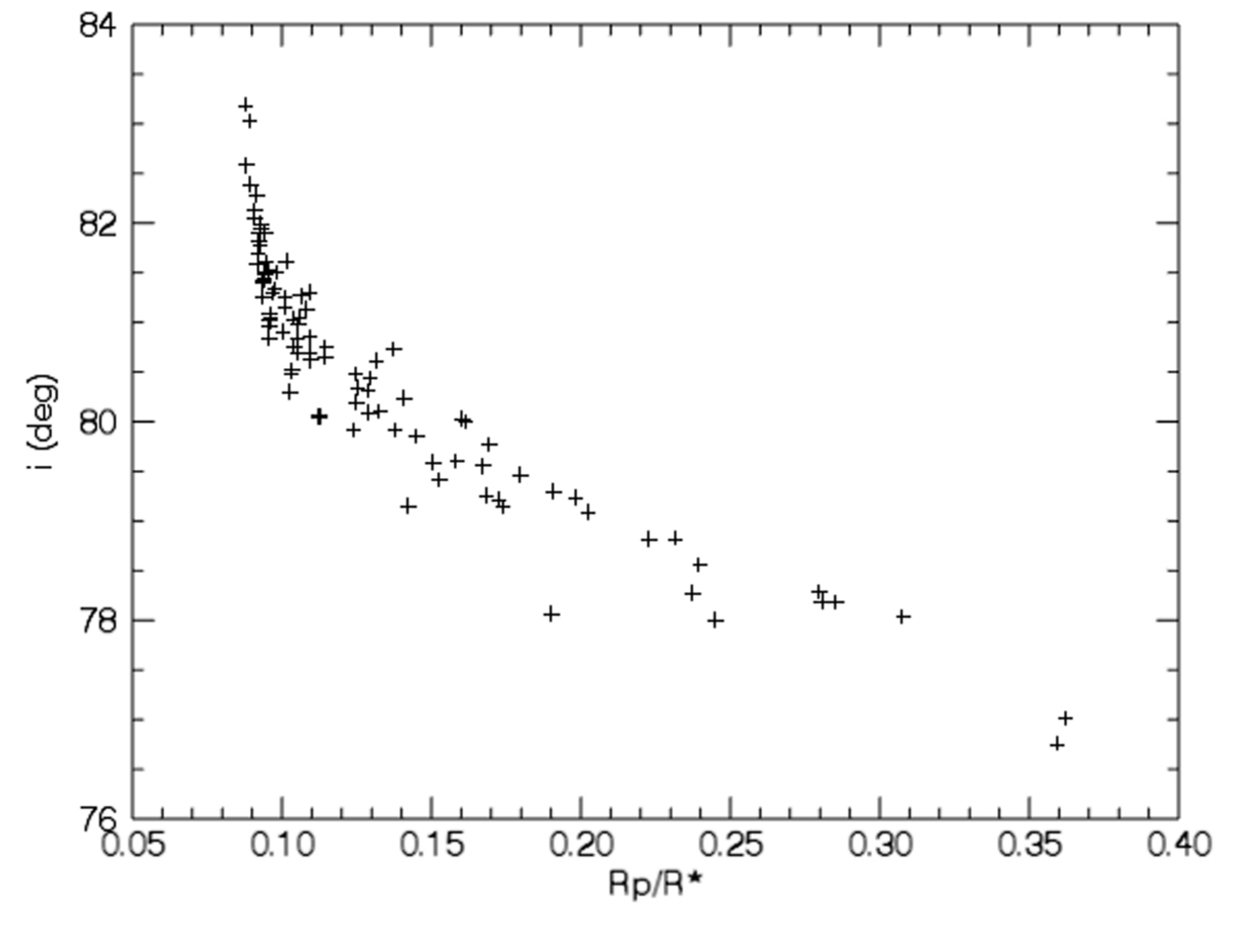}}
\caption{Example of the different results of the MCMC simulations as derived using \texttt{UTM/UFIT} \citep{deeg_2014}. Possible solutions for inclination $i$ and relative planetary radius $R_\mathrm{p}/R_{\star}$ are plotted. The strong correlation between these parameters -- generally not reflected in the error bars of MCMC simulations -- is clearly visible.}
\label{corr}
\end{center}
\end{figure}

Figure \ref{corr} shows the unresolved correlation between inclination $i$ and relative radius $R_\mathrm{p}/R_*$. Assuming that planetary systems are randomly oriented, their inclination distribution follows a probability density proportional to $\sin i$, which can be assumed to be uniform in the inclination range of $80.7-79.1\degr$ that corresponds to the planet's orbital inclination. Given the absence of any further indicator that prefers some planet-size anywhere within the range $0.7-1.4\,R_\mathrm{Jup}$, we consider that the planet has a radius within that range with a uniform distribution of probability. Indications of the same problem were reported for Kepler-447b \citep{Lillo-Box_2015}, who decided to increase the error bar of the planetary radius.

\begin{table}[t]
%\centering
\caption{Orbital and planetary parameters.}
\label{System-Param}
%\scalebox{0.9}{
\begin{tabular}{lr}
\hline
\hline
\noalign{\smallskip}
Period [day]                                          & $1.257850 \pm 0.000002$	 \\
\noalign{\smallskip}
Mid-transit epoch [BJD-2454833 day]                   & $2358.70889 \pm 0.00024$ \\
\noalign{\smallskip}
Transit duration [day]                                & 0.0409$\pm$0.0002        \\
\noalign{\smallskip}
Linear limb darkening coefficient                     & $0.514_{-0.138}^{+0.169}$\\
\noalign{\smallskip}
Quadratic limb darkening coefficient                  & $0.290_{-0.128}^{+0.173}$\\       
\noalign{\smallskip}
Scaled semi-major axis $a_\mathrm{p}/R_{\star}$       & $6.05 \pm 0.50$          \\
\noalign{\smallskip}
Impact parameter $b$                                  &  $0.9-1.05$              \\
\noalign{\smallskip}
Orbit inclination [deg]                               & $80.7-79.1$              \\
\noalign{\smallskip}
Eccentricity                                          & 0 (fixed)                \\
\noalign{\smallskip}
Planet-to-star radius ratio $R_\mathrm{p}/R_{\star}$  & $0.09-0.18$              \\
\noalign{\smallskip}
RV semi-amplitude variation [\kms]                    & 0.3509$\pm$0.0010        \\
\noalign{\smallskip}
Systemic velocity (FIES) [\kms]                       & $-$4.9434$\pm$0.0026     \\
\noalign{\smallskip}                                                              
Systemic velocity (HARPS)[\kms]                       & $-$4.7511$\pm$0.0011     \\
\noalign{\smallskip}                                                   
Semi major axis [AU]               & $0.0220\pm 0.0018$        \\
\noalign{\smallskip}
Planet mass   $M_\mathrm{p}$ [\Mjup]                  & 1.774$\pm$0.079          \\
\noalign{\smallskip}
Planet radius $R_\mathrm{p}$ [\Rjup] & $ 0.71-1.41$             \\  
\noalign{\smallskip}
\hline
%\hline
\end{tabular}
%\tablenotetext{\tablenotemark{a}{Values are calculated using the stellar radius.}}
\end{table}

We can determine the effective temperature of the planet using the host star's parameters and considering that the planet is tidally locked. Assuming a low albedo and that the stellar flux energy is distributed over the star-facing hemisphere results in an equilibrium temperature of the planet of 1750~K. In the Kepler bandpass, reflected light is therefore expected to dominate thermal emission from the planet when observed near secondary eclipse. A search for such eclipses did not lead to any reliable detection. We found an amplitude upper limit of 76~ppm, which implies an upper limit to the planet-to-star surface-brightness ratio of 0.01, corresponding to a planetary geometric albedo of less than 0.4. The planet is therefore rather dark, as are most hot Jupiters \citep{cowan_2011}. We note that this albedo-limit is not affected by the uncertainties in the planet's size.

\section{Summary and final remarks}

K2-31b is the first hot Jupiter detected and confirmed by the K2 mission. K2-31 is a relatively bright (V=10.8~mag) G7\,V star of $M_\star$=0.91$\pm$0.06~\Msun\ and $R_\star$=0.78$\pm$0.07~\Rsun, hosting a short period (\pper=1.26 days) Jupiter-like planet with a mass of 1.8~\Mjup. \citet{Dai_2016} published additional RV measurements of K2-31. Their results of the stellar and planetary mass are consistent with our results. Their values for the stellar radius are slightly higher than our values. This would also change the absolute planetary radius but the error is dominated by the uncertainty of the relative planetary radius.
The radius of the planet could only be determined in the range of $0.7-1.4$~\Rjup\ because of the grazing transit combined with the K2 low acquisition rate. The detection of correlations between different parameters is not guaranteed when using MCMC simulations. The use of Monte Carlo simulations would suggest more precise results for the planetary radius with error bars that are unrealistically small. The estimates of the planetary radius and orbital parameters could be improved with additional optical and near-infrared transit observations, as recently shown by \citet{Mancini2014} for the grazing transit of \object{WASP-67}. 

Blended eclipsing binary systems whose light is diluted by the main target inside the photometric mask often cause V-shaped transits that are difficult to distinguish from rare grazing planetary transits \citep{Brown2003}. It is possible that some V-shaped transits are discarded because they are misidentified as background binaries. In the case of K2-31, the V-shaped transit is caused by a grazing hot-Jupiter. \citet{oshagh_2015} stated that the number of known grazing transiting exoplanets are lower than what we expect. This new detection is important to understand the detection biases. Follow-up observations are needed in order to assess the true nature of V-shaped transit signals in the absence of higher time-sampling and/or multi-color transit photometry. The short orbital period (1.26\,days) and transit duration ($\sim$1\,hour) are a definitive advantage for the observation schedule and only a small number of short period hot Jupiters are detected to date.

\vspace{2pt}

Based on observations made with the Nordic Optical Telescope (NOT) operated on the island of La Palma by Nordic Optical Telescope Scientific Association in the Observatorio del Roque de Los Muchachos and with the Telescopio Carlos Sanchez (TCS) at Teide Observatory on Tenerife, both of the Instituto de Astrof\'\i sica de Canarias. HD and DN acknowledge support by grant AYA2012-39346-C02-02 of the Spanish Secretary of State for R\&D\&i (MINECO). This paper includes data collected by the Kepler-K2 mission. Funding for the Kepler mission is provided by the NASA Science Mission directorate.
We would like to thank the anonymous referee for the valuable comments which improved the paper.

%\vspace{6pt}

\facilities{Kepler (K2), NOT (FIES), ESO-3.6m (HARPS), Sanchez (FAST-CAM).}

\software{\texttt{DST}, \texttt{EXOTRANS}, \texttt{SME}, \texttt{TAP}, \texttt{EXOFAST}, \texttt{TLCM}, \texttt{PyTransit}, \texttt{UTM/UFIT}}


\listofchanges

\end{document}